%% file: charged2014.tex
\documentclass{PoS}
\pdfoutput=1
\usepackage{graphicx}
\usepackage{subfig}
\usepackage{float}
\usepackage{amsmath}
\usepackage{amssymb}
\usepackage{amsthm}
\usepackage{latexsym}
\usepackage{mciteplus}
\usepackage{enumitem}

\newcommand{\newc}{\newcommand*}
\newc{\hobs}{\ensuremath{H_\text{obs}}}
\newc{\mhpm}{\ensuremath{m_{\hpm}}}
\newc{\hpm}{\ensuremath{H^{\pm}}}
\newc{\mhsm}{\ensuremath{ m_{h_{\mathrm{SM}}}}}
\newc{\hsm}{\ensuremath{h_{\rm SM}}}            
\newc{\gev}{\ensuremath{\,\mathrm{GeV}}}
\newc{\tev}{\ensuremath{\,\mathrm{TeV}}}
\newc{\GeV}{\gev} 
\newc{\TeV}{\tev}

\def\etmiss{{E\hskip -0.2cm\slash \hskip 0.01cm }_T}

\title{Search for Charged Higgs bosons via decays to $W^\pm$ and a 125 GeV Higgs at the Large Hadron Collider}

\ShortTitle{Search for $H^\pm\to W^\pm \hobs$ at the LHC}

\author{Rikard Enberg$^a$, \speaker{William Klemm}$^a$, Stefano Moretti$^{b}$, Shoaib Munir$^a$ and~Glenn~Wouda$^a$\\
        \llap{$^a$}Department of Physics and Astronomy, Uppsala University\\ Box 516 , SE-Uppsala, Sweden\\
        \llap{$^b$}School of Physics \& Astronomy, University of Southampton, Highfield,\\ Southampton SO17 1BJ, UK
        
        E-mail: \email{rikard.enberg@physics.uu.se}, \email{william.klemm@physics.uu.se}, \email{S.Moretti@soton.ac.uk}, \email{shoaib.munir@physics.uu.se}, \email{glenn.wouda@physics.uu.se}}

\abstract{The recent observation of a 125 GeV neutral Higgs boson ($\hobs$) provides additional input for charged Higgs boson searches in the $H^\pm \to W^\pm \hobs$ decay channel at the Large Hadron Collider (LHC).  We reassess the discovery potential in this channel, which is important for $\hpm$ heavier than the top quark mass.  When $\hobs$ decays to a $b\bar{b}$ pair, knowledge of the Higgs mass aids in the kinematic selection of signal events. We perform a signal-to-background analysis to demonstrate the LHC prospects for charged Higgs discovery in the resulting channel $pp\to t(\bar{b})H^-\to \ell^\pm\nu_\ell jj bb\bar{b}(\bar{b})$+h.c. for standard (300 fb$^{-1}$) and high (3000 fb$^{-1}$) luminosities at design energy, $\sqrt{s}=14\tev$. We find that regions of the parameter space of several two-Higgs doublet models, consistent with constraints from LHC Higgs searches and $b$-physics observables, are testable in this channel.}

\FullConference{Prospects for Charged Higgs Discovery at Colliders\\
                 16-18 September 2014\\
                 Uppsala University Sweden}

\begin{document}

\section{Introduction}

The observation of a Higgs boson ($\hobs$)~\cite{Aad:2012tfa,*Chatrchyan:2012ufa} at the Large Hadron Collider (LHC)  may be just the first glimpse into the rich phenomenology of a larger Higgs sector. Indeed, many models require additional Higgs states, including charged Higgs bosons ($\hpm$), the observation of which would be a clear sign of physics beyond the standard model (SM). Experimental searches for $\hpm$ have largely focused on decays to $tb$ or $\tau\nu$, which dominate much of the parameter space of many models.  However, when kinematically allowed, the decay $\hpm\to W^\pm \hobs$ can become significant for many parameter configurations. Earlier studies~\cite{Drees:1999sb,*Moretti:2000yg} demonstrated the potential of this channel, and the newfound knowledge of the Higgs mass, $m_{\hobs}\approx 125\gev$, provides an additional input for the analysis and a constraint on extensions of the Higgs sector (see also~\cite{Coleppa:2014cca} for a recent discussion). Here we describe a collider analysis for the $\hpm\to W^\pm \hobs$ channel and determine its sensitivity at the LHC, which we compare to the possible signal strengths of several two Higgs doublet models (2HDMs) compatible with current experimental observations.

\section{Collider Analysis}

The main production channel at the LHC for a charged Higgs above the top mass is typically $pp\to t(b)H^\pm$,\footnote{This should be interpreted as $pp\to t(\bar{b})H^{-}+pp\to \bar{t}(b)H^{+}$. Throughout this text, we will not distinguish fermions and anti-fermions when their identity is unspecified and/or can be inferred.} which is possible through the coupling of the charged Higgs to third generation quarks.  
Focusing then on $H^\pm\to\hobs W^\pm$ decays, we consider the subsequent decay $\hobs\to b\bar{b}$, as both $b$-quarks are observable, allowing us to directly reconstruct the observed 125\gev state, and because SM-like Higgs bosons in this mass range decay dominantly in this channel.\footnote{In principle, other decay channels could also be competitive. It has been suggested that, especially in analyses dominated by systematic errors, the $\hobs\to\tau^+\tau^-$ channel could be useful despite additional missing energy from $\tau$ decays and a reduced branching ratio, largely as a result of lower backgrounds~\cite{Coleppa:2014cca}.}
The process we then wish to search for is $pp\to (b)tH^\pm \to (b)b W^\mp W^\pm \hobs \to (b)bbb jj \ell\nu_\ell$, where one of the $W$-bosons (from either $H^\pm$ or top decay) decays leptonically and the other hadronically.  The presence of a single lepton allows us to avoid multi-jet backgrounds, while requiring one hadronic $W$ avoids additional unseen neutrinos, making the event reconstruction more straightforward.  The main background for this process is $t\bar{t}b(\bar{b})$, where either an additional $b$-tagged jet combines with a $b$-jet from a top decay or an additional $b\bar{b}$ pair mimics an $\hobs\to b\bar{b}$ decay.

To get a measure of the sensitivity that could be obtained at the $14\tev$ LHC, we generate the $t(b)\hpm$ signal using Pythia 6.4.28~\cite{Sjostrand:2006za} with the MATCHIG~\cite{Alwall:2004xw} add-on to avoid double counting among $bg\to t\hpm$ and $gg\to tb\hpm$ processes, and all $t(b)WX,X\to b\bar{b}$ backgrounds with MadGraph5~\cite{Alwall:2011uj}. Both signal and background undergo parton showering and hadronization using Pythia 8~\cite{Sjostrand:2007gs} and are further processed with the DELPHES 3~\cite{deFavereau:2013fsa} detector simulation using experimental parameters based on the ATLAS experiment with modified $b$-tagging efficiencies.\footnote{The $b$-tagging efficiency chosen is $\epsilon_\eta\tanh(0.03 p_T - 0.4)$, with $\epsilon_\eta = 0.7$ for central ($|\eta|\leq 1.2$) and $\epsilon_\eta = 0.6$ for forward ($1.2\leq|\eta|\leq 2.5$) jets, and the transverse momentum, $p_T$, in GeV. This is a conservative choice compared with high-luminosity projections.}
 To reconstruct our signal events and reduce background, we use the following procedure, inspired by previous studies~\cite{Drees:1999sb,*Moretti:2000yg}, with an additional top veto:
\begin{enumerate}[noitemsep]
\item \textbf{Event selection:} Require events to have at least 3 $b$-tagged jets, at least 2 light jets, one lepton ($e/\mu$), and missing energy $\etmiss\ge 20\gev$. All objects must have transverse momentum $p_T\ge 20\gev$ and rapidity $|\eta|\leq 2.5$, with separation $\Delta R \ge 0.4$ from other objects.
\item \textbf{Hadronic ${W}$ reconstruction:} Choose the pair of light jets with invariant mass $m_{jj}$ closest to $m_W$, and reject the event if no pair satisfies $|m_{jj}-m_W|\leq 30\gev$.
\item \textbf{Leptonic $W$ reconstruction:} Attributing all $\etmiss$ to a neutrino from a $W$ decay, use the observed lepton to find the longitudinal component of the neutrino momentum, $p_{\nu,z}$ by imposing the mass constraint $m_{\ell\nu}=m_W$. The solution will have a twofold ambiguity as a result of the quadratic nature of the constraint. For two real solutions, keep both.  For complex solutions, discard the imaginary component and retain a single real $p_{\nu,z}$.
\refstepcounter{enumi}
\item[(\number\value{enumi}.)]\label{it:topveto}\textbf{Top veto (high mass region, ``veto first''):} If two top quarks can be reconstructed from reconstructed $W$'s and any unassigned jets, with both satisfying $|m_{Wj}-m_t|\leq 20\gev$, reject the event. The jets used may or may not be $b$-tagged. 
\item \textbf{$\hobs$ reconstruction:} Choose the pair of $b$-tagged jets with invariant mass $m_{bb}$ closest to $m_{\hobs}\sim 125\gev$, and reject the event if no pair satisfies $|m_{bb}-m_{\hobs}|\leq 15\gev$.
\item[\refstepcounter{enumi}(\number\value{enumi}.)] \textbf{Top veto (low mass region, ``veto second''):} Same as (\ref{it:topveto}.), but $b$-jets used in $\hobs$ reconstruction are excluded.
\item \textbf{Top reconstruction:} From the reconstructed $W$'s and remaining $b$-tagged jets, identify the best top quark candidate, determined by the $Wb$ combination with the invariant mass $m_{Wb}$ closest to $m_t$. If the selected combination includes one leptonic $W$ solution, discard the other. If there is no good candidate with $|m_{Wb}-m_t|\leq 30\gev$, reject the event.
\item\textbf{$\hpm$ reconstruction}: Combine the reconstructed $\hobs$ with the remaining $W$ to yield the discriminating variable $m_{W\hobs}$. If there are two leptonic $W$'s remaining, retain both values of $m_{W\hobs}$.
\end{enumerate}

\begin{figure}[ht!]
\centering
\subfloat[]{
\label{fig:sigback-a}
\includegraphics[angle=0,width=0.5\textwidth]{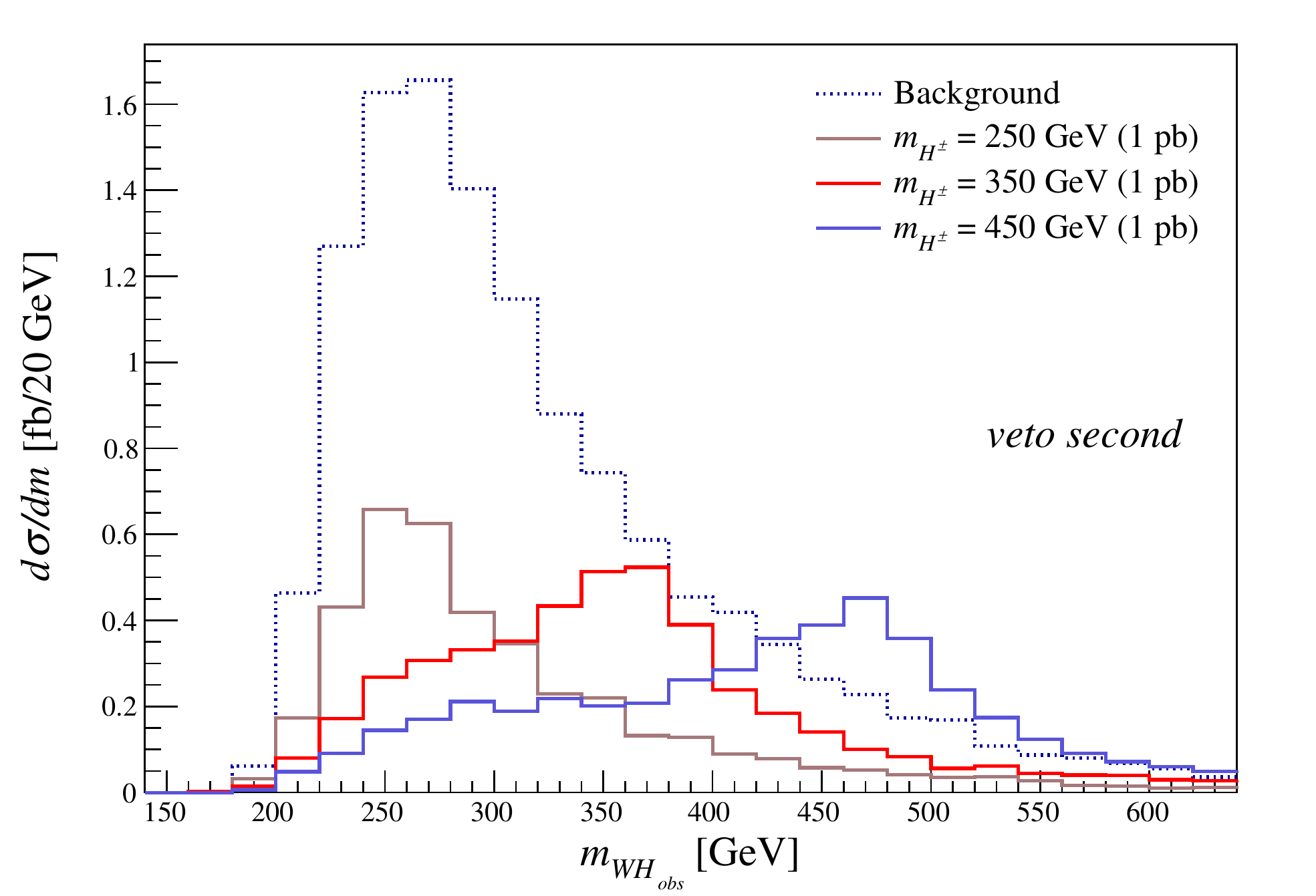}
}
\subfloat[]{
\label{fig:sigback-b}
\includegraphics[angle=0,width=0.5\textwidth]{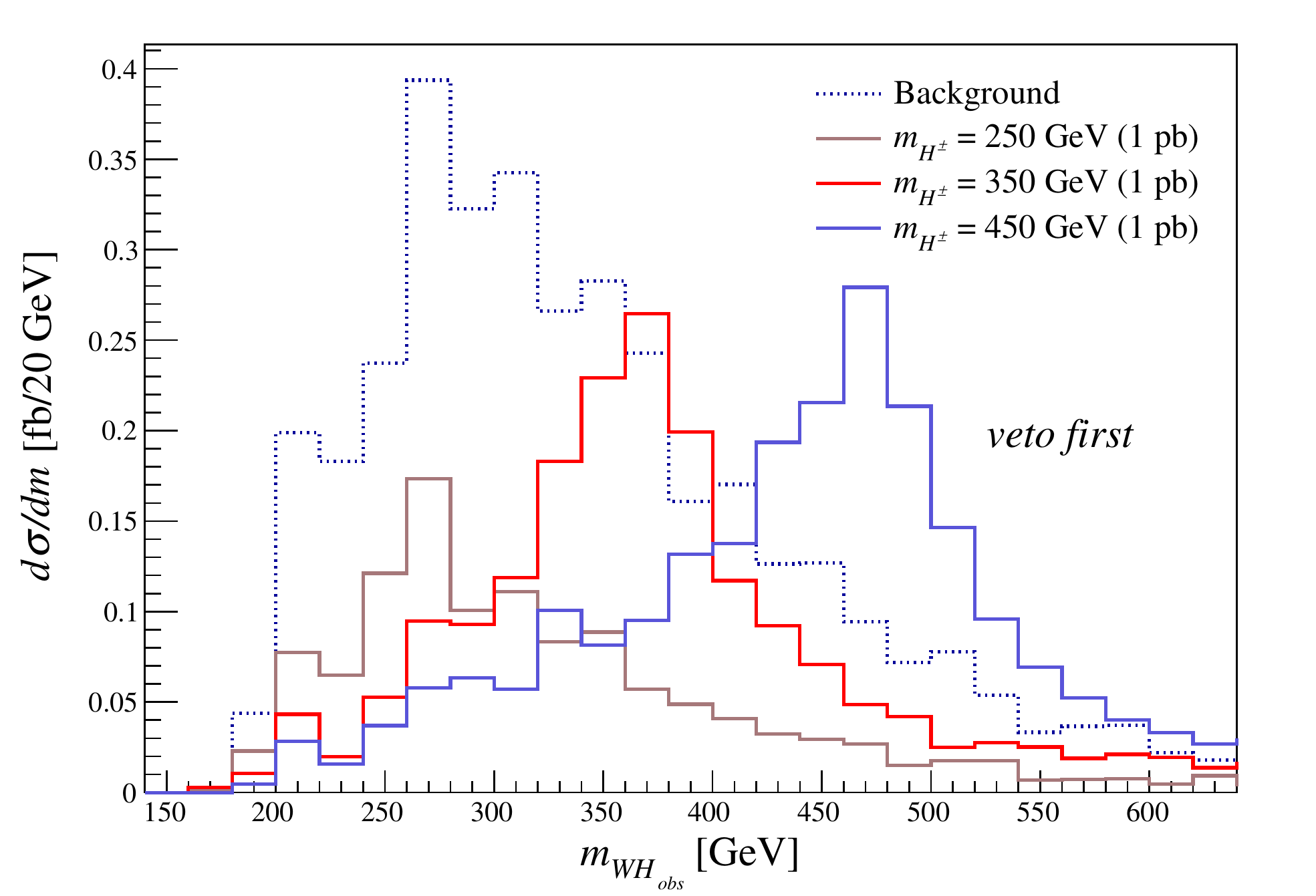}
}
\caption{Reconstructed signal and background $m_{W\hobs}$ with two different top vetoes (described in text). The signals are normalized to $\sigma(pp\rightarrow tH^\pm)\times BR(H^\pm\rightarrow h W^\pm) \times BR(h\rightarrow b\bar{b})=1$\,pb before selection and cuts.}
\label{fig:sigback}
\end{figure}

The background is often able to mimic the signal by combining a $b$-jet from a top decay with an additional $b$-tagged jet to reconstruct the $\hobs$.  In order to remove this type of event, the top veto should be applied prior to the $\hobs$ reconstruction (``veto first'').  However, because of the relative sizes of the masses involved, for charged Higgs masses not too far above the $\hpm\to\hobs W^\pm$ threshold, one of the resulting $b$-jets combines with the $W^\pm$ to give an invariant mass $m_{bW}\approx m_t$ in a large fraction of the available phase space. Such signal events are cut in the ``veto first'' scenario, negating the benefits of the background reduction. For lower mass searches, we then postpone the top veto until after the $\hobs$ reconstruction (``veto second''). In practice, we consider both top vetoes for a given mass choose the one which maximizes the statistical signal, $S/\sqrt{B}$, and find that ``veto second'' is preferable for $\mhpm \lesssim 350\gev$.\footnote{A full experimental analysis considering all sources of error may place greater emphasis on background reduction, which would likely shift this value.} This is apparent in Fig.~\ref{fig:sigback}, where the $\mhpm$ resonant peak is also evident. To further improve significance, for each $\mhpm$ we consider, we place a cut on the range of reconstructed $m_{W\hobs}$ which maximizes $S/\sqrt{B}$.

\section{Models}

One of the most straightforward extensions of the Higgs sector is the 2HDM, in which there are two scalar electroweak doublets, $\Phi_1$ and $\Phi_2$, which can each in general acquire a vacuum expecation value $v_i$ and couple to each other and standard model particles. After symmetry breaking, the Higgs sector in a 2HDM contains five states: two $CP$-even ($h,H$), one pseudoscalar ($A$), and two charged ($\hpm$). In general, either $h$ or $H$ could correspond to $\hobs$. Here we will focus on results for the case where $\hobs$ is the lighter state, $h$.

The Yukawa couplings to fermions are \textit{a priori} free parameters of the theory but can easily lead to large tree-level flavor-changing neutral currents (FCNCs). One way to suppress FCNCs is to introduce a $Z_2$-symmetry which only allows each type of fermion to couple to a single doublet~\cite{Glashow:1976nt,*Paschos:1976ay}.  There are four possible $Z_2$ assignments, and here we will consider two cases: Type I (2HDM-I), where fermions only couple to $\Phi_2$; and Type II (2HDM-II), where down-type quarks and leptons couple to $\Phi_1$ and up-type quarks couple to $\Phi_2$. In these models, the Yukawa couplings are determined entirely by the parameter $\tan\beta=v_2/v_1$. Another mechanism for controlling FCNCs is to require that the two Higgs doublets have Yukawa matrices which are proportional to one another, or aligned. Here we consider the case where all fermions couple to both $\Phi_1$ and $\Phi_2$ with aligned couplings, known as the A2HDM~\cite{Pich:2009sp}.

In order to see whether the $\hpm\to W^\pm \hobs$ channel is a useful probe of these models, we scan their parameter spaces for regions with a strong signal.  We require that the lightest $CP$-even Higgs have a mass consistent with the observed state, $123\leq m_h \leq 127 \gev$, and that the heaver $H$ be non-degenerate, $135\leq m_H\leq 500\gev$.  To satisfy electroweak constraints, we require $m_A=\mhpm$, and we consider 
$\hpm$ masses in the region above the $W^\pm h$ threshold, $200\leq \mhpm\leq 500\gev$. For the 2HDM-II, this is modified to $320\leq \mhpm\leq 500\gev$ to reflect $b$-physics constraints.  In Type-I and II, we consider  
$1.5\leq\tan\beta\leq 6$, where the branching ratio $BR(\hpm\to W^\pm h)$ is typically largest.\footnote{For a full description of the parameter scans, and results for $\hobs~=~H$ and supersymmetric models, see~\cite{Enberg:2014pua}.}

Some of the strongest constraints on 2HDMs come from $b$-physics observables, and we subject the scans to 95\% confidence limits on $BR(\bar{B}\to X_s\gamma)$, $BR(B_u\to \tau\nu)$, and $BR(B_s\to \mu^+\mu^-)$ given in~\cite{superiso}, and on $\Delta M_{B_d}$ from~\cite{Mahmoudi:2009zx}. For $Z_2$-symmetric models, the parameter space scanned was chosen to satisfy these constraints, as described in~\cite{Mahmoudi:2009zx}, and for the A2HDM, $b$-physics observables were calculated with SuperIso-v3.4~\cite{superiso}. In addition, we subject all Higgs states other than $h$ to LEP, Tevatron, and LHC constraints using HiggsBounds-v4.1.3~\cite{Bechtle:2013wla}.  Finally, we consider signal strength $\mu^X$ of $\hobs$ decay channels which have been recently measured, where $\mu^X=\sigma(pp\to\hobs\to X)/\sigma(pp\to \hsm\to X)$, with a $125\gev$ SM Higgs boson $\hsm$. We determine the theoretical counterparts of $\mu^X$ with HiggsSignals-v1.20~\cite{Bechtle:2013xfa} for $X= \gamma\gamma,\,ZZ$ and compare with the measurements of $\mu^{\gamma \gamma} = 1.13 \pm 0.24$, $\mu^{ZZ} = 1.0 \pm 0.29$ by CMS~\cite{CMS-PAS-HIG-14-009}.

\section{Results}
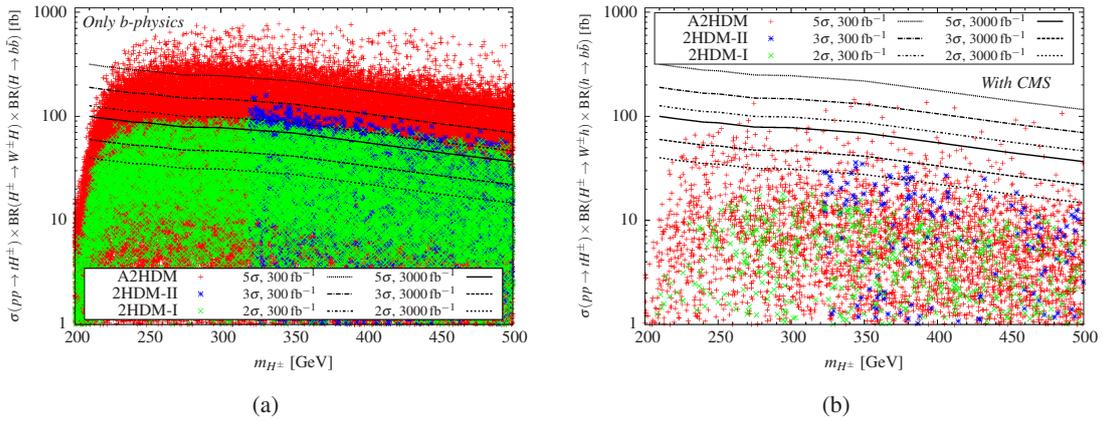
\begin{figure}[ht!]
\centering
\centering
\subfloat[]{%
	\label{fig:results-a}%
	\resizebox{0.49\textwidth}{!}{\input{./ALL2hdm_mhpm_XS.tex}}
}
\subfloat[]{%
	\label{fig:results-b}%
	\resizebox{0.49\textwidth}{!}{\input{./CMS2hdm_mhpm_XS.tex}}
}
\caption{Signal strength $\sigma(pp\rightarrow tH^\pm)\times BR(H^\pm\rightarrow h W^\pm)\times BR(h\rightarrow b\bar{b})$ from 2HDM scans described in text, along with expected statistical sensitivity contours $S/\sqrt{B}=2,3,5$, for an integrated luminosity of $\mathcal{L}=300$\,fb$^{-1}$ at the next LHC run and $\mathcal{L}=3000$\,fb$^{-1}$ at the High Luminosity LHC, both at $\sqrt{s}=14$\tev. Results are shown (a) without and (b) with CMS constraints on $\mu^{\gamma \gamma}$ and $\mu^{ZZ}$. }
\label{fig:results}
\end{figure}

Fig.~\ref{fig:results} shows the results of the parameter scans along with the sensitivity expected from the collider analysis.  For the $Z_2$-symmetric 2HDMs, we find a large number of points which are potentially discoverable at a high-luminosity LHC. However, both of these models see deviations of $h$ from $\hsm$ for the points with the largest signal and consequently show less detection potential when the very SM-like CMS constraints are imposed. The A2HDM shows even stronger signals, well within reach of even the standard luminosity LHC. The effect of the CMS constraints is again severe, but some points still remain testable at lower luminosities. The $\hpm\to W^\pm\hobs$ channel can be a useful probe of 2HDMs at the LHC, particularly at high luminosities.

\acknowledgments{
This work was in part funded by the Swedish Research Council under contracts 2007-4071 and 621-2011-5107.
The work of S.~Moretti has been funded in part through the NExT Institute. 
The computational work was in part carried out on
resources provided by the Swedish National Infrastructure
for Computing (SNIC) at Uppsala Multidisciplinary Center
for Advanced Computational Science (UPPMAX) under Projects p2013257 and SNIC 2014/1-5.
}

\input{charged2014.bbl}

\end{document}

%% file: ALL2hdm_mhpm_XS.tex
% GNUPLOT: LaTeX picture with Postscript
\begingroup
  \makeatletter
  \providecommand\color[2][]{%
    \GenericError{(gnuplot) \space\space\space\@spaces}{%
      Package color not loaded in conjunction with
      terminal option `colourtext'%
    }{See the gnuplot documentation for explanation.%
    }{Either use 'blacktext' in gnuplot or load the package
      color.sty in LaTeX.}%
    \renewcommand\color[2][]{}%
  }%
  \providecommand\includegraphics[2][]{%
    \GenericError{(gnuplot) \space\space\space\@spaces}{%
      Package graphicx or graphics not loaded%
    }{See the gnuplot documentation for explanation.%
    }{The gnuplot epslatex terminal needs graphicx.sty or graphics.sty.}%
    \renewcommand\includegraphics[2][]{}%
  }%
  \providecommand\rotatebox[2]{#2}%
  \@ifundefined{ifGPcolor}{%
    \newif\ifGPcolor
    \GPcolortrue
  }{}%
  \@ifundefined{ifGPblacktext}{%
    \newif\ifGPblacktext
    \GPblacktexttrue
  }{}%
  % define a \g@addto@macro without @ in the name:
  \let\gplgaddtomacro\g@addto@macro
  % define empty templates for all commands taking text:
  \gdef\gplbacktext{}%
  \gdef\gplfronttext{}%
  \makeatother
  \ifGPblacktext
    % no textcolor at all
    \def\colorrgb#1{}%
    \def\colorgray#1{}%
  \else
    % gray or color?
    \ifGPcolor
      \def\colorrgb#1{\color[rgb]{#1}}%
      \def\colorgray#1{\color[gray]{#1}}%
      \expandafter\def\csname LTw\endcsname{\color{white}}%
      \expandafter\def\csname LTb\endcsname{\color{black}}%
      \expandafter\def\csname LTa\endcsname{\color{black}}%
      \expandafter\def\csname LT0\endcsname{\color[rgb]{1,0,0}}%
      \expandafter\def\csname LT1\endcsname{\color[rgb]{0,1,0}}%
      \expandafter\def\csname LT2\endcsname{\color[rgb]{0,0,1}}%
      \expandafter\def\csname LT3\endcsname{\color[rgb]{1,0,1}}%
      \expandafter\def\csname LT4\endcsname{\color[rgb]{0,1,1}}%
      \expandafter\def\csname LT5\endcsname{\color[rgb]{1,1,0}}%
      \expandafter\def\csname LT6\endcsname{\color[rgb]{0,0,0}}%
      \expandafter\def\csname LT7\endcsname{\color[rgb]{1,0.3,0}}%
      \expandafter\def\csname LT8\endcsname{\color[rgb]{0.5,0.5,0.5}}%
    \else
      % gray
      \def\colorrgb#1{\color{black}}%
      \def\colorgray#1{\color[gray]{#1}}%
      \expandafter\def\csname LTw\endcsname{\color{white}}%
      \expandafter\def\csname LTb\endcsname{\color{black}}%
      \expandafter\def\csname LTa\endcsname{\color{black}}%
      \expandafter\def\csname LT0\endcsname{\color{black}}%
      \expandafter\def\csname LT1\endcsname{\color{black}}%
      \expandafter\def\csname LT2\endcsname{\color{black}}%
      \expandafter\def\csname LT3\endcsname{\color{black}}%
      \expandafter\def\csname LT4\endcsname{\color{black}}%
      \expandafter\def\csname LT5\endcsname{\color{black}}%
      \expandafter\def\csname LT6\endcsname{\color{black}}%
      \expandafter\def\csname LT7\endcsname{\color{black}}%
      \expandafter\def\csname LT8\endcsname{\color{black}}%
    \fi
  \fi
  \setlength{\unitlength}{0.0500bp}%
  \begin{picture}(7200.00,5040.00)%
%    \gplgaddtomacro\gplbacktext{%
%      \csname LTb\endcsname%
%    }%
    \gplgaddtomacro\gplfronttext{%
      \csname LTb\endcsname%
      \put(2522,1317){\makebox(0,0)[r]{\strut{}A2HDM}}%
      \csname LTb\endcsname%
      \put(2522,1097){\makebox(0,0)[r]{\strut{}2HDM-II}}%
      \csname LTb\endcsname%
      \put(2522,877){\makebox(0,0)[r]{\strut{}2HDM-I}}%
      \csname LTb\endcsname%
      \put(1104,704){\makebox(0,0)[r]{\strut{} 1}}%
      \put(1104,2043){\makebox(0,0)[r]{\strut{} 10}}%
      \put(1104,3381){\makebox(0,0)[r]{\strut{} 100}}%
      \put(1104,4720){\makebox(0,0)[r]{\strut{} 1000}}%
      \put(1210,528){\makebox(0,0){\strut{} 200}}%
      \put(2142,528){\makebox(0,0){\strut{} 250}}%
      \put(3074,528){\makebox(0,0){\strut{} 300}}%
      \put(4007,528){\makebox(0,0){\strut{} 350}}%
      \put(4939,528){\makebox(0,0){\strut{} 400}}%
      \put(5871,528){\makebox(0,0){\strut{} 450}}%
      \put(6803,528){\makebox(0,0){\strut{} 500}}%
      \put(426,2739){\rotatebox{-270}{\makebox(0,0){\strut{}\footnotesize{$\sigma(pp \rightarrow tH^{\pm}) \times {\rm BR}(H^\pm \rightarrow W^\pm H)\times {\rm BR}(H\rightarrow b\bar{b})$ [fb]}}}}%
      \put(4006,204){\makebox(0,0){\strut{}$m_{H^\pm}$ [GeV]}}%
      \put(1303,4590){\makebox(0,0)[l]{\strut{}{\it Only b-physics}}}%
      \put(4301,1317){\makebox(0,0)[r]{\strut{}\footnotesize{5$\sigma$,\;300\,fb$^{-1}$}}}%
      \csname LTb\endcsname%
      \put(4301,1097){\makebox(0,0)[r]{\strut{}\footnotesize{3$\sigma$,\;300\,fb$^{-1}$}}}%
      \csname LTb\endcsname%
      \put(4301,877){\makebox(0,0)[r]{\strut{}\footnotesize{2$\sigma$,\;300\,fb$^{-1}$}}}%
      \csname LTb\endcsname%
      \put(6080,1317){\makebox(0,0)[r]{\strut{}\footnotesize{5$\sigma$,\;3000\,fb$^{-1}$}}}%
      \csname LTb\endcsname%
      \put(6080,1097){\makebox(0,0)[r]{\strut{}\footnotesize{3$\sigma$,\;3000\,fb$^{-1}$}}}%
      \csname LTb\endcsname%
      \put(6080,877){\makebox(0,0)[r]{\strut{}\footnotesize{2$\sigma$,\;3000\,fb$^{-1}$}}}%
    }%
    \gplbacktext
    \put(0,0){\includegraphics{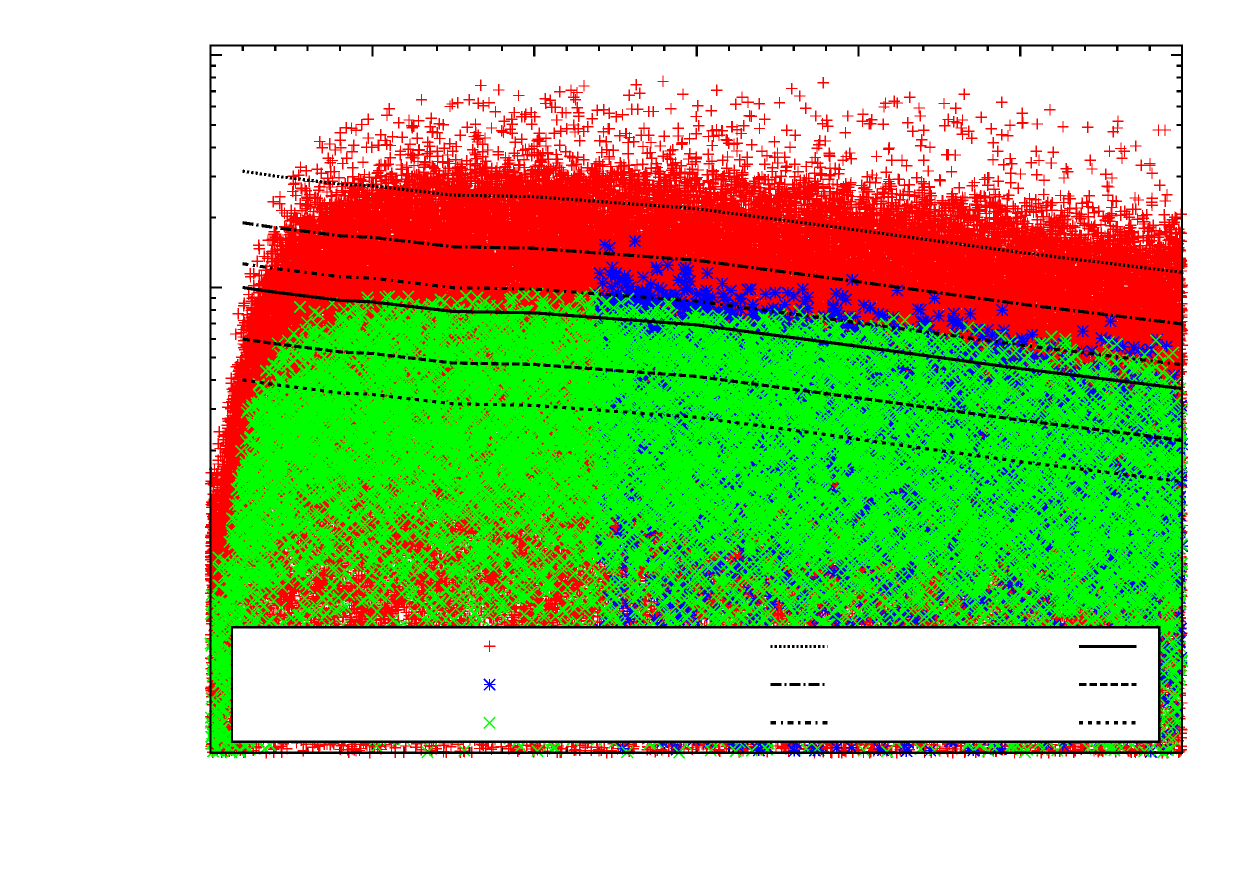}}%
    \gplfronttext
  \end{picture}%
\endgroup

%% file: CMS2hdm_mhpm_XS.tex
% GNUPLOT: LaTeX picture with Postscript
\begingroup
  \makeatletter
  \providecommand\color[2][]{%
    \GenericError{(gnuplot) \space\space\space\@spaces}{%
      Package color not loaded in conjunction with
      terminal option `colourtext'%
    }{See the gnuplot documentation for explanation.%
    }{Either use 'blacktext' in gnuplot or load the package
      color.sty in LaTeX.}%
    \renewcommand\color[2][]{}%
  }%
  \providecommand\includegraphics[2][]{%
    \GenericError{(gnuplot) \space\space\space\@spaces}{%
      Package graphicx or graphics not loaded%
    }{See the gnuplot documentation for explanation.%
    }{The gnuplot epslatex terminal needs graphicx.sty or graphics.sty.}%
    \renewcommand\includegraphics[2][]{}%
  }%
  \providecommand\rotatebox[2]{#2}%
  \@ifundefined{ifGPcolor}{%
    \newif\ifGPcolor
    \GPcolortrue
  }{}%
  \@ifundefined{ifGPblacktext}{%
    \newif\ifGPblacktext
    \GPblacktexttrue
  }{}%
  % define a \g@addto@macro without @ in the name:
  \let\gplgaddtomacro\g@addto@macro
  % define empty templates for all commands taking text:
  \gdef\gplbacktext{}%
  \gdef\gplfronttext{}%
  \makeatother
  \ifGPblacktext
    % no textcolor at all
    \def\colorrgb#1{}%
    \def\colorgray#1{}%
  \else
    % gray or color?
    \ifGPcolor
      \def\colorrgb#1{\color[rgb]{#1}}%
      \def\colorgray#1{\color[gray]{#1}}%
      \expandafter\def\csname LTw\endcsname{\color{white}}%
      \expandafter\def\csname LTb\endcsname{\color{black}}%
      \expandafter\def\csname LTa\endcsname{\color{black}}%
      \expandafter\def\csname LT0\endcsname{\color[rgb]{1,0,0}}%
      \expandafter\def\csname LT1\endcsname{\color[rgb]{0,1,0}}%
      \expandafter\def\csname LT2\endcsname{\color[rgb]{0,0,1}}%
      \expandafter\def\csname LT3\endcsname{\color[rgb]{1,0,1}}%
      \expandafter\def\csname LT4\endcsname{\color[rgb]{0,1,1}}%
      \expandafter\def\csname LT5\endcsname{\color[rgb]{1,1,0}}%
      \expandafter\def\csname LT6\endcsname{\color[rgb]{0,0,0}}%
      \expandafter\def\csname LT7\endcsname{\color[rgb]{1,0.3,0}}%
      \expandafter\def\csname LT8\endcsname{\color[rgb]{0.5,0.5,0.5}}%
    \else
      % gray
      \def\colorrgb#1{\color{black}}%
      \def\colorgray#1{\color[gray]{#1}}%
      \expandafter\def\csname LTw\endcsname{\color{white}}%
      \expandafter\def\csname LTb\endcsname{\color{black}}%
      \expandafter\def\csname LTa\endcsname{\color{black}}%
      \expandafter\def\csname LT0\endcsname{\color{black}}%
      \expandafter\def\csname LT1\endcsname{\color{black}}%
      \expandafter\def\csname LT2\endcsname{\color{black}}%
      \expandafter\def\csname LT3\endcsname{\color{black}}%
      \expandafter\def\csname LT4\endcsname{\color{black}}%
      \expandafter\def\csname LT5\endcsname{\color{black}}%
      \expandafter\def\csname LT6\endcsname{\color{black}}%
      \expandafter\def\csname LT7\endcsname{\color{black}}%
      \expandafter\def\csname LT8\endcsname{\color{black}}%
    \fi
  \fi
  \setlength{\unitlength}{0.0500bp}%
  \begin{picture}(7200.00,5040.00)%
%    \gplgaddtomacro\gplbacktext{%
%      \csname LTb\endcsname%
%    }%
    \gplgaddtomacro\gplfronttext{%
      \csname LTb\endcsname%
      \put(2522,4602){\makebox(0,0)[r]{\strut{}A2HDM}}%
      \csname LTb\endcsname%
      \put(2522,4382){\makebox(0,0)[r]{\strut{}2HDM-II}}%
      \csname LTb\endcsname%
      \put(2522,4162){\makebox(0,0)[r]{\strut{}2HDM-I}}%
      \csname LTb\endcsname%
      \put(1104,704){\makebox(0,0)[r]{\strut{} 1}}%
      \put(1104,2043){\makebox(0,0)[r]{\strut{} 10}}%
      \put(1104,3381){\makebox(0,0)[r]{\strut{} 100}}%
      \put(1104,4720){\makebox(0,0)[r]{\strut{} 1000}}%
      \put(1210,528){\makebox(0,0){\strut{} 200}}%
      \put(2142,528){\makebox(0,0){\strut{} 250}}%
      \put(3074,528){\makebox(0,0){\strut{} 300}}%
      \put(4007,528){\makebox(0,0){\strut{} 350}}%
      \put(4939,528){\makebox(0,0){\strut{} 400}}%
      \put(5871,528){\makebox(0,0){\strut{} 450}}%
      \put(6803,528){\makebox(0,0){\strut{} 500}}%
      \put(426,2739){\rotatebox{-270}{\makebox(0,0){\strut{}\footnotesize{$\sigma(pp \rightarrow tH^{\pm}) \times {\rm BR}(H^\pm \rightarrow W^\pm h)\times {\rm BR}(h\rightarrow b\bar{b})$ [fb]}}}}%
      \put(4006,204){\makebox(0,0){\strut{}$m_{H^\pm}$ [GeV]}}%
      \put(5500,3800){\makebox(0,0)[l]{\strut{}{\it With CMS}}}%
      \put(4301,4602){\makebox(0,0)[r]{\strut{}\footnotesize{5$\sigma$,\;300\,fb$^{-1}$}}}%
      \csname LTb\endcsname%
      \put(4301,4382){\makebox(0,0)[r]{\strut{}\footnotesize{3$\sigma$,\;300\,fb$^{-1}$}}}%
      \csname LTb\endcsname%
      \put(4301,4162){\makebox(0,0)[r]{\strut{}\footnotesize{2$\sigma$,\;300\,fb$^{-1}$}}}%
      \csname LTb\endcsname%
      \put(6080,4602){\makebox(0,0)[r]{\strut{}\footnotesize{5$\sigma$,\;3000\,fb$^{-1}$}}}%
      \csname LTb\endcsname%
      \put(6080,4382){\makebox(0,0)[r]{\strut{}\footnotesize{3$\sigma$,\;3000\,fb$^{-1}$}}}%
      \csname LTb\endcsname%
      \put(6080,4162){\makebox(0,0)[r]{\strut{}\footnotesize{2$\sigma$,\;3000\,fb$^{-1}$}}}%
    }%
    \gplbacktext
    \put(0,0){\includegraphics{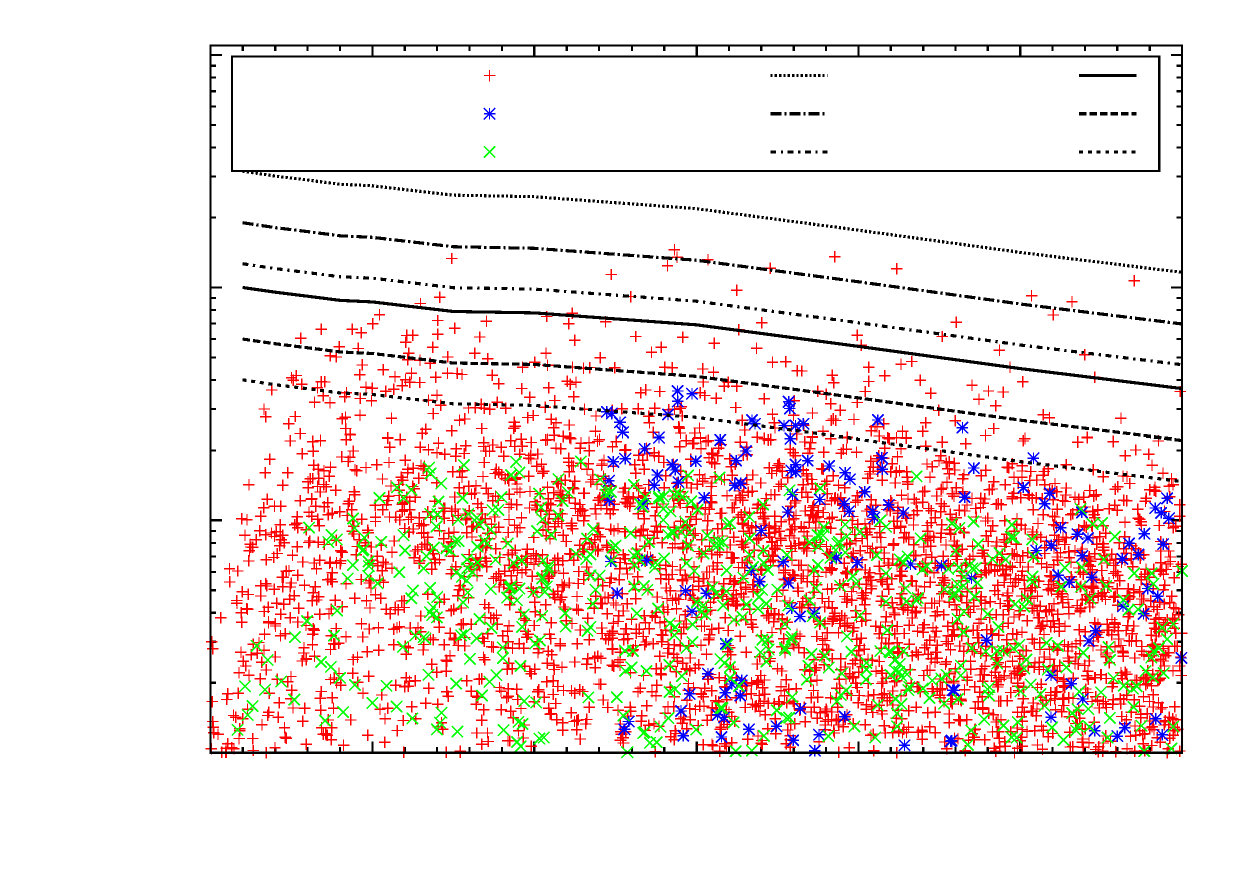}}%
    \gplfronttext
  \end{picture}%
\endgroup

%% file: charged2014.bbl
\ifx\mcitethebibliography\mciteundefinedmacro
\PackageError{unsrtM.bst}{mciteplus.sty has not been loaded}
{This bibstyle requires the use of the mciteplus package.}\fi